\documentclass[thmsa]{article}
\usepackage{sw20lart}

\input{tcilatex}
\begin{document}

\title{Towards a Statistical Geometrodynamics }
\author{Ariel Caticha \\
{\small Department of Physics, University at Albany--SUNY, Albany, NY 12222,
USA}}
\date{}
\maketitle

\begin{abstract}
Can the spatial distance between two identical particles be explained in
terms of the extent that one can be distinguished from the other? Is the
geometry of space a macroscopic manifestation of an underlying microscopic
statistical structure? Is geometrodynamics derivable from general principles
of inductive inference? Tentative answers are suggested by a model of
geometrodynamics based on the statistical concepts of entropy, information
geometry, and entropic dynamics.
\end{abstract}

\section{\strut Introduction}

The purpose of dynamical theories is to predict or explain the changes
observed in physical systems on the basis of information that is codified
into what one calls the states of the system. One common view is that these
dynamical theories -- the laws of physics -- are successful because they
happen to reflect the true laws of nature.

Here I wish to follow an alternative path: perhaps once the relevant
information has been identified the question of predicting changes is just a
matter of careful consistent manipulation of the available information. If
this turns out to be the case, then the laws of physics should follow
directly from rules for processing information, that is, the rules of
probability theory \cite{Cox46} and the method of maximum entropy (ME) \cite
{Jaynes57}--\cite{Skilling88}.\footnote{%
On terminology: The ME method is designed for processing information to
update from a prior probability distribution to a posterior distribution.
(The terms `prior' and `posterior' are used with similar meanings in the
context of Bayes' theorem.) The ME\ method is usually understood in the
restricted sense that one updates from a prior distribution that happens to
be uniform -- this is the usual postulate of equal a priori probabilities.
Here we adopt a broader meaning that includes updates from arbitrary priors
and which involves the maximization of \emph{relative} entropy. Since all
entropies are relative to some prior, be it uniform or not, the qualifier
`relative' is redundant and will henceforth be omitted. For a brief account
of the ME method in a form that is convenient for our current purposes see 
\cite{Caticha01a}.}

There are some indications that this point of view is worth pursuing.
Indeed, thermodynamics is a prime example of a fundamental physical theory
that can be derived from general principles of inference \cite{Jaynes57}.
Quantum mechanics provides a second, less trivial, and less well known
example \cite{Caticha98}. Both theories follow from a correct specification
of the subject matter, that is, an appropriate choice of variables -- this
is the truly difficult step -- plus probabilistic and entropic arguments.

A third independent clue is found when one attempts to derive classical
dynamical theories from purely entropic arguments. The surprising outcome is
that the resulting ``entropic'' dynamics (ED) shows remarkable similarities
with the general theory of relativity -- geometrodynamics (GD). The general
purpose of this paper is to take the first tentative steps towards
explaining geometrodynamics as a form of entropic dynamics.

The procedure to derive an ED involves three steps \cite{Caticha02}. The
first step is to identify the subject matter and the corresponding space of
observable states or, perhaps more appropriately, the space of macrostates.
This is not easy because there exists no systematic way to search for the
right macrovariables; it is a matter of taste and intuition, trial and error.

The second step is to define a quantitative measure of the change or the
``distance'' from one state to another. Although in general the choice of
distance is not unique an exception occurs when the macrostates can be
interpreted as probability distributions over some appropriate space of
microstates. Then there is a natural distance which is given by the
Fisher-Rao information metric \cite{Fisher25}\cite{Rao45} (its uniqueness is
discussed in \cite{Cencov81}\cite{Campbell86}; for a brief heuristic
derivation see \cite{Caticha01b}). It measures the extent to which one
probability distribution can be distinguished from another. This second step
-- assigning a statistical distance -- is not straightforward either: more
inspired guesswork is needed unless the right microstates happen to be known
beforehand.\footnote{%
The recognition that spaces of probability distributions are metric spaces
has nevertheless been fruitful in statistics, where the subject is known as
Information Geometry \cite{Amari85}\cite{Rodriguez02}, and in physics \cite
{Weinhold75}--\cite{Balian86}.}

The third and final step is easier. We ask: Given the initial and the final
states, what trajectory is the system expected to follow? The question
implicitly assumes that there is a trajectory, that in moving from one state
to another the system will pass through a continuous set of intermediate
states, and that information about the initial and final states is
sufficient to determine them. The answer follows from a principle of
inference, the ME principle, and not from any additional ``physical''
postulates.

The resulting ED is elegant and not trivial: the system moves along a
geodesic but the geometry of the space of states is curved and possibly
quite complicated. Since the only available clock is the system itself there
is no reference to an external physical time. The natural intrinsic time is
defined by the change of the system itself -- in ED time is change -- and
can only be obtained after the equations of motion are solved. ED is a
timeless Machian dynamics and its features resemble those advocated by
Barbour \cite{Barbour94}: it is reversible; it can be derived from a Jacobi
action principle rather than the more familiar action principle of Hamilton;
and its canonical Hamiltonian formulation is an example of a dynamics driven
by constraints.

The similarities to GD are striking. For example, in GD there is no
reference to an external physical time. The proper time interval along any
curve between an initial and a final three-dimensional geometries of space
is determined only after solving the Einstein equations of motion \cite
{Baierlein62}. The absence of an external time has been a serious impediment
in understanding GD because it is not clear which variables represent the
true gravitational degrees of freedom \cite{York79}--\cite{Isham93}. GD is
also derived from a Jacobi action principle \cite{Brown89}\cite{Hartle88}
and its canonical Hamiltonian formulation is an example of a dynamics driven
by constraints \cite{Dirac58}--\cite{Hojman76}. The question, therefore, is
whether GD is an example of ED. The answer requires identifying those
variables that describe the true degrees of freedom of the gravitational
field.

The tentative steps of making assumptions about the subject matter, the
macrostates, and about how to associate a probability distribution to each
of them are taken in section 2. We want to predict the evolution of the
three-dimensional geometry of space. The problem is that space is invisible.
What we see is not space, but matter in space and we do not quite know how
to disentangle which properties should be attributed to the matter and which
to space. The best one can do is to choose the simplest form of matter: a
substance that is neutral to all interactions and is itself describable by a
minimal number of attributes. This ideal form of matter is a dust of
identical particles; being neutral they will only interact gravitationally,
and being identical the issue of what it is that distinguishes them -- size,
mass, flavor -- does not arise. Thus we assume there is nothing to space
beyond what can be learned from observing the evolving distribution of dust
particles. The geometry of space is just the geometry of all the distances
between dust particles. Furthermore, we assume this geometry is of
statistical origin. Identical particles that are close together are easy to
confuse, those that are far apart are easy to distinguish. The distance
between two neighboring particles is the distinguishability distance given
by a Fisher-Rao metric. Notice that the Fisher-Rao metric is used in two
conceptually different ways. One is to distinguish successive states of the
same system, the other is to distinguish different neighboring particles.
The first is related to time, the second to space.

Having decided what system is under study and how it is statistically
described we can proceed to define its ED. In section 3, as a warm up
problem, we develop the ED of a single point, and then, in section 4, we
generalize to the whole dust cloud. Although the resulting statistical GD is
not Einstein's GD of space-time -- an indication that the states and
variables we have chosen do not accurately describe the gravitational
degrees of freedom -- it is close enough to be encouraging. The model GD
developed here corresponds to what is called an ultralocal or strong gravity
theory \cite{Isham76}--\cite{Pilati82}. We do not recover the notion of
space-time but we do find an embryonic form of Lorentz invariance in that
simultaneity is relative. Finally, in section 5 we summarize our conclusions.

\section{The Geometry of a Dust Cloud}

Consider a cloud of identical specks of dust suspended in an otherwise empty
space. And there is nothing else; in particular, there are no rulers and no
clocks, just dust. Our goal is to study how the cloud evolves. We do this by
keeping track of individual specks of dust.

Being identical the particles are easy to confuse. The only distinction
between two of them is that one happens to be here while the other is over
there. To distinguish one speck of dust from another we assign labels or
coordinates to each particle. We assume that three real numbers $(y^1,y^2,$ $%
y^3)$ are sufficient.

Of course, particles can be mislabeled. Then the ``true'' coordinates $y$
are unknown and one can only provide an estimate, $x$. Let $p(y|x)dy$ be the
probability that the particle labeled $x$ should have been labeled $y$. The
labels $x$ are introduced to distinguish one particle from another, but can
we distinguish a particle at $x$ from another at $x+dx$? If $dx$ is small
enough the corresponding probability distributions $p(y|x)$ and $p(y|x+dx)$
overlap considerably and it is easy to confuse them. We seek a quantitative
measure of the extent to which these two distributions can be distinguished.

The following crude argument is intuitively appealing. Consider the relative
difference, 
\begin{equation}
\frac{p(y|x+dx)-p(y|x)}{p(y|x)}=\frac{\partial \log p(y|x)}{\partial x^{{}i}}%
\,dx^{{}i}.
\end{equation}
The expected value of this relative difference does not provide us with the
desired measure of distinguishability: it vanishes identically. However, the
variance does not vanish, 
\begin{equation}
d\lambda ^2=\int d^3y\,p(y|x)\,\frac{\partial \log p(y|x)}{\partial x^{{}i}}%
\,\frac{\partial \log p(y|x)}{\partial x^{{}j}}\,dx^{{}i}dx^{{}j}\stackrel{%
\limfunc{def}}{=}\gamma _{ij}\,(x)dx^{{}i}dx^{{}j}\,\,.
\label{Fisher metric}
\end{equation}
This is the measure of distinguishability we seek. Except for an overall
multiplicative constant, the Fisher-Rao metric $\gamma _{ij}$ is the only
Riemannian metric that adequately reflects the underlying statistical nature
of the abstract manifold of the distributions $p(y|x)$ \cite{Cencov81}\cite
{Campbell86}.

We take the further step of interpreting $d\lambda $ as the \emph{spatial}
distance of the three-dimensional space the dust inhabits. Indeed, one would
normally say that the reason it is easy to confuse two particles is that
they happen to be too close together. We argue in the opposite direction and 
\emph{explain} that the reason the particles at $x$ and at $x+dx$ are close
together is \emph{because} they are difficult to distinguish.

The origin of the uncertainty will be left unspecified; perhaps it is due to
a limit on the ultimate resolution of observation devices, or perhaps, as
with a particle undergoing Brownian motion, the uncertainty might be caused
by a fluctuating physical agent. It is required, however, that two particles
at the same location in space must be affected by the same uncertainty, the
same irreducible noise. Then the noise is not linked to the particle, but to
the place, and we might as well say that the source of the irreducible noise
is space itself. This is somewhat analogous to the principle of equivalence:
it is the fact that all particles irrespective of their mass move along the
same trajectories in a gravitational field that allows us to eliminate the
notion of a gravitational field and attribute their common behavior to a
single universal agent, the curvature of space-time.

To assign an explicit $p(y|x)$ and explore the geometry it induces we will
consider what is perhaps the simplest possibility. We assume that the
uncertainty in the coordinate $x$ is small so that $p(y|x)$ is sharply
localized in a neighborhood about $x$ and within this very small region
curvature effects can be neglected. Further, we assume that particles are
labeled by the expected values $\langle y^i\rangle =x^i$ and that the
information that happens to be necessary for the purpose of prediction of
future behavior is given by the second moments $\langle
(y^i-x^i)(y^j-x^j)\rangle =C^{ij}(x)$. This is physically reasonable: for
each particle we have estimates for its position and of the small margin of
error. Then $p(y|x)$ can be determined maximizing entropy relative to an
appropriate prior. To the extent that curvature effects are negligible, the
underlying space is flat and translationally invariant. Thus, symmetry
suggests a uniform prior and the resulting ME distribution is Gaussian, 
\begin{equation}
p(y|x)=\frac{C^{1/2}}{(2\pi )^{3/2}}\,\exp \left[ -\frac
12C_{ij}(y^i-x^i)(y^j-x^j)\right] ,
\end{equation}
where $C_{ij}$ is the inverse of the covariance coefficients $C^{ij}$, $%
C^{ik}C_{kj}=\delta _j^i$, and $C\equiv \det C_{ij}$. The corresponding
metric is obtained substituting into eq.(\ref{Fisher metric}). For small
uncertainties $C_{ij}(x)$ is constant within the region where $p(y|x)$ is
appreciable and the result is 
\begin{equation}
\gamma _{ij}(x)=C_{ij}(x)\,.
\end{equation}
The metric changes smoothly over space and, in general, space is curved. The
connection, the curvature, and other aspects of its Riemannian geometry can
be computed in the standard way. The probability distributions, 

\begin{equation}
p(y|x)=\frac{\gamma ^{1/2}(x)}{(2\pi )^{3/2}}\,\exp \left[ -\frac 12\gamma
_{ij}(x)(y^i-x^i)(y^j-x^j)\right] \,,  \label{p(y|x)}
\end{equation}
also vary smoothly with $x$. 

To summarize, we have succeeded in describing the information geometry that
derives from considerations of distinguishability among particles. The idea
is rather general but was developed explicitly only for the special case of
small uncertainties, that is, for particles that can be localized within
regions much smaller than those where curvature effects become appreciable.
An interesting question that will not be addressed here concerns the
extension to those situations of extreme curvature found near singularities.

Before discussing dynamics we mention that there is one very peculiar
feature of the distance $d\lambda $, eq.(\ref{Fisher metric}), that may be
very significant: $d\lambda ^2$ is dimensionless. The metric $\gamma
_{ij}(x) $ allows one to measure spatial lengths in terms of a local
standard, the local uncertainty width. This immediately raises the question
of how to compare the uncertainty widths, and therefore lengths, at two
distant locations. One possibility, which we pursue in the rest of this
paper, is that $\gamma _{ij}$ describes the Riemannian geometry of space.
This amounts to asserting that the uncertainty widths are the same
everywhere, they provide us with a universal standard of length. A second,
more intriguing possibility, which we will explore elsewhere, is that all
the information metric $\gamma _{ij}$ allows us to do is to compare the
lengths of small segments in different orientations at the same location; it
allows one to measure angles. Then $\gamma _{ij}$ does not describe the
geometry of space completely, it only describes its conformal geometry.

\section{Entropic Dynamics of a Single Point}

In this section we develop the ED of a single Gaussian distribution, an
analogue of GD in zero spatial dimensions. Let $\Gamma $ be the space of
states. The points in $\Gamma $ are Gaussian distributions with zero mean $%
\langle y\rangle =0$, 
\begin{equation}
p(y|\gamma )=\frac{\gamma ^{1/2}}{(2\pi )^{3/2}}\,\exp \left( -\frac
12\gamma _{ij}y^iy^j\right) \,,
\end{equation}
where $\gamma =\det \gamma _{ij}$ and $y=(y^1,y^2,y^3)$ are points in $R^3$.
Whether $\gamma $ denotes the matrix $\gamma _{ij}$ or its determinant
should, in what follows, be clear from the context. Since $\gamma
_{ij}=\gamma _{ji}$ is symmetric, $\Gamma $ is a six dimensional space. 

The following notation is convenient: the derivative $\partial /\partial
\gamma _{ij}$ of a function $F(\gamma )$ is defined so that $dF$ takes the
simple form 
\begin{equation}
dF\stackrel{\limfunc{def}}{=}\frac{\partial F}{\partial \gamma _{ij}}d\gamma
_{ij}\,.
\end{equation}
$\partial F/\partial \gamma _{ij}$ coincides with the usual partial
derivative times $(1+\delta _{ij})/2$. To operate with $\partial /\partial
\gamma _{ij}$ we only need to find out how it acts on $\gamma _{kl}$ and on
its inverse $\gamma ^{kl}$. We find 
\begin{equation}
\frac{\partial \gamma _{kl}}{\partial \gamma _{ij}}=\frac 12\left( \delta
_k^i\delta _l^j+\delta _l^i\delta _k^j\right) \stackrel{\limfunc{def}}{=}%
\delta _{kl}^{ij}\,\,\,\text{and\thinspace \thinspace }\,\frac{\partial
\gamma ^{kl}}{\partial \gamma _{ij}}=-\frac 12\left( \gamma ^{ki}\gamma
^{lj}+\gamma ^{kj}\gamma ^{li}\right) .
\end{equation}
Note that $\delta _{kl}^{ij}\gamma _{ij}=\gamma _{kl}$ and $\delta
_{kl}^{ij}\gamma ^{kl}=\gamma ^{ij}$. We will also need to differentiate the
determinant $\gamma =\det \gamma _{ij}$, 
\begin{equation}
d\gamma =\gamma \gamma ^{ij}d\gamma _{ij}\quad \text{or}\quad \frac{\partial
\gamma }{\partial \gamma _{ij}}=\gamma \gamma ^{ij}\,.
\end{equation}

The Fisher-Rao metric $g^{ij\,kl}$ on the space $\Gamma $ is 
\begin{equation}
g^{ij\,kl}=\int dy\,p(y|\gamma )\frac{\partial \log p(y|\gamma )}{\partial
\gamma _{ij}}\frac{\partial \log p(y|\gamma )}{\partial \gamma _{kl}}=\frac
14\left( \gamma ^{ki}\gamma ^{lj}+\gamma ^{kj}\gamma ^{li}\right) \,,
\label{metric g}
\end{equation}
and its inverse metric, defined by $g^{ij\,kl}g_{kl\,mn}=\delta _{mn}^{ij}$%
\thinspace , is 
\begin{equation}
g_{kl\,mn}=\gamma _{km}\gamma _{ln}+\gamma _{kn}\gamma _{lm}\,.
\end{equation}

Now we can tackle the dynamics. The key to the question ``Given initial and
final states, what trajectory is the system expected to follow?'' lies in
the implicit assumption that there exists a continuous trajectory. This
means that large changes are the result of a continuous succession of very
many small changes; the problem of studying large changes is reduced to the
simpler problem of studying small changes.

We want to determine the states along a short segment of the trajectory as
the system moves from an initial state $\gamma $ to a neighboring final
state $\gamma +\Delta \gamma $. To find the intermediate states we reason
that in going from the initial to the final state the system must pass
through a halfway point, that is, an intermediate state that is equidistant
from $\gamma $ and $\gamma +\Delta \gamma $. Finding the halfway point
clearly determines the trajectory: first find the halfway point, and use it
to determine `quarter of the way' points, and so on.\thinspace But there is
nothing special about halfway states. In general, we can assert that the
system must pass through intermediate states $\gamma _{{}{}{}\omega }$ such
that, having already moved a distance $d\ell $ away from the initial $\gamma 
$, there remains a distance $\omega d\ell $ to be covered to reach the final 
$\gamma +\Delta \gamma $; $\omega $ is any positive number. 

The basic dynamical question can be rephrased as follows: The system is
initially described by the probability distribution $p(y|\gamma )$ and we
are given the new information that the system has moved to one of the
neighboring states in the family $p(y|\gamma _{{}\omega })$. Which $%
p(y|\gamma _{{}\omega })$ do we select? Phrased in this way it is clear that
this is precisely the kind of problem to be tackled using the ME method.%
\footnotemark[1]  The selected distribution is that which maximizes the
relative entropy of $p(y|\gamma _{{}\omega })$ relative to a prior
distribution $p_{\limfunc{old}}$. Since in the absence of new information
there is no reason to change one's mind, when there are no constraints the
selected posterior distribution should coincide with the prior distribution.
Therefore the prior $p_{\text{old}}$ is the initial state $p(y|\gamma )$.
Thus, to determine the intermediate state $\gamma _{{}\omega }=\gamma
+d\gamma $ one varies over $d\gamma _{ij}$ to maximize 
\begin{eqnarray}
S\left[ p(y|\gamma _{{}\omega }),p(y|\gamma )\right] &=&-\int dy\,p(y|\gamma
+d\gamma )\log \frac{p(y|\gamma +d\gamma )}{p(y|\gamma )}  \nonumber \\
&=&-\frac 12g^{ij\,kl}d\gamma _{ij}d\gamma _{kl}=-\frac 12d\ell ^2\,,
\end{eqnarray}
subject to the constraint $d\ell _f=\omega d\ell $ where 
\begin{equation}
d\ell _f^2=g^{ij\,kl}\left( \Delta \gamma _{ij}-d\gamma _{ij}\right) \left(
\Delta \gamma _{kl}-d\gamma _{kl}\right) \,.
\end{equation}
Introducing a Lagrange multiplier $\lambda /2$, 
\begin{equation}
0=\delta \left[ -\frac 12g^{ij\,kl}d\gamma _{ij}d\gamma _{kl}+\frac \lambda
2\left( \omega ^2d\ell ^2-d\ell _f^2\right) \right] \,,
\end{equation}
then, the selected $d\gamma _{ij}$ is given by 
\begin{equation}
d\gamma _{ij}=\chi \Delta \gamma _{ij}\quad \text{where}\quad \chi =\frac
\lambda {1+\lambda (1-\omega ^2)}\,.  \label{single point dgamma}
\end{equation}
Substituting $d\gamma _{ij}$ into $d\ell $ and $d\ell _f$ we get $d\ell
=\chi \Delta \ell $ and $d\ell _f=(1-\chi )\Delta \ell \,$, so that $\chi
=(1+\omega )^{-1}$ with $0<\chi <1$ and 
\begin{equation}
d\ell +d\ell _f=\Delta \ell \,.
\end{equation}
The interpretation is clear: the three states $\gamma $, $\gamma _{{}\omega
} $ and $\gamma +\Delta \gamma $ lie on a straight line. The expected
trajectory is the geodesic that passes through the given initial and final
states.

Note that each different value of $\omega $ provides a different criterion
to select the trajectory and an inconsistency would arise if these criteria
led to different trajectories. It is reassuring to find that indeed the ED
trajectory is independent of the value $\omega $.

ED determines the vector tangent to the trajectory $d\gamma /d\ell $, but
not the actual velocity $d\gamma /dt$. In conventional forms of dynamics the
distance $\ell $ along the trajectory is related to an external time $t$
through a Hamiltonian which fixes the evolution relative to external clocks.
But here the only clock available is the system itself which can only
provide an internal, intrinsic time. It is best to define the intrinsic time
so that motion looks simple. A natural definition consists in stipulating
that the system moves with unit velocity, then the intrinsic time is given
by the distance $\ell $ itself. The intrinsic time interval is the amount of
change. A peculiar feature of this notion of time is that intervals are not
a priori known, they are determined only after the equations of motion are
solved and the actual trajectory is determined.

The geodesics in the space $\Gamma $ are obtained minimizing the Jacobi
action 
\begin{equation}
J[\gamma ]=\int_{\eta _i}^{\eta _f}d\eta \,L(\gamma ,\dot{\gamma})\,,
\label{single point J}
\end{equation}
where $\eta $ is an arbitrary parameter along the trajectory and $\dot{\gamma%
}_{ij}=d\gamma _{ij}/d\eta $. The Lagrangian is just the arc length 
\begin{equation}
L(\gamma ,\dot{\gamma})=\left( g^{ij\,kl}\dot{\gamma}_{ij}\dot{\gamma}%
_{kl}\right) ^{1/2}=\left( \frac 12\gamma ^{ik}\gamma ^{jl}\dot{\gamma}_{ij}%
\dot{\gamma}_{kl}\right) ^{1/2}.
\end{equation}

The canonical momenta are 
\begin{equation}
\pi ^{mn}=\frac{\partial L}{\partial \dot{\gamma}_{mn}}=\frac 1{2L}\gamma
^{ik}\gamma ^{jl}\dot{\gamma}_{ij}\delta _{kl}^{mn}=\frac 1{2L}\gamma
^{mi}\gamma ^{nj}\dot{\gamma}_{ij}\,,  \label{pi momenta}
\end{equation}
and have a fixed magnitude 
\begin{equation}
g_{ij\,kl}\pi ^{ij}\pi ^{kl}=1\,.  \label{constraint pi}
\end{equation}
The canonical Hamiltonian vanishes identically, 
\begin{equation}
H_{\text{can}}(\gamma ,\pi )=\dot{\gamma}_{ij}\pi ^{ij}-L(\gamma ,\dot{\gamma%
})\equiv 0\,,
\end{equation}
because the Lagrangian is homogeneous of first degree in the velocities. The
manifest reparametrization invariance of the action $J[\gamma ]$
conveniently reflects the absence of an external time with respect to which
the system could possibly evolve.

Since variations of the momenta are constrained to preserve their magnitude
the action principle is 
\begin{equation}
I[\gamma ,\pi ,N]=\int_{\eta _i}^{\eta _f}d\eta \left[ \dot{\gamma}_{ij}\pi
^{ij}-N(\eta )h(\gamma ,\pi )\right] \,,
\end{equation}
where 
\begin{equation}
h(\gamma ,\pi )\stackrel{\limfunc{def}}{=}\frac 12g_{ij\,kl}\pi ^{ij}\pi
^{kl}-\frac 12\,,
\end{equation}
and $N(\eta )$ are Lagrange multipliers that at each instant $\eta $ enforce
the constraints 
\begin{equation}
h(\gamma ,\pi )=0\,.
\end{equation}

Equations of motion are obtained varying with respect to $\gamma $ and $\pi $
with $\gamma $ fixed at the endpoints $\delta \gamma _{ij}(\eta _i)=\delta
\gamma _{ij}(\eta _f)=0$. Then 
\begin{eqnarray}
\dot{\gamma}_{mn} &=&N\frac{\partial h}{\partial \pi ^{mn}}=2N\gamma
_{mi}\gamma _{nj}\pi ^{ij}  \label{single point motion a} \\
\dot{\pi}^{mn} &=&-N\frac{\partial h}{\partial \gamma _{mn}}=-2N\gamma
_{ij}\pi ^{mi}\pi ^{nj}.  \label{single point motion b}
\end{eqnarray}
There is no equation of motion for $N$. Comparing eq.(\ref{pi momenta}) and (%
\ref{single point motion a}) we get 
\begin{equation}
N(\eta )=L(\gamma ,\dot{\gamma})=\frac{d\ell }{d\eta }\,,
\end{equation}
which is recognized as the ``lapse'' function which gives the increase of
intrinsic time $\ell $ per unit increase of the parameter $\eta $. Then the
equations of motion simplify to 
\begin{eqnarray}
\frac{d\gamma _{mn}}{d\ell } &=&\frac{\partial h}{\partial \pi ^{mn}}%
=2\gamma _{mi}\gamma _{nj}\pi ^{ij}  \label{single point motion c} \\
\frac{d\pi ^{mn}}{d\ell } &=&-\frac{\partial h}{\partial \gamma _{mn}}%
=-2\gamma _{ij}\pi ^{mi}\pi ^{nj}.  \label{single point motion d}
\end{eqnarray}

One can check that $dh/d\eta =0$. Therefore if $h=0$ initially, the
constraint will be consistently preserved by the evolution. One can also
check that the action $I[\gamma ,\pi ,N]$ is invariant under the gauge
transformations 
\begin{equation}
\delta \gamma _{mn}=\varepsilon (\eta )\frac{\partial h}{\partial \pi ^{mn}}%
,\quad \delta \pi ^{mn}=-\varepsilon (\eta )\frac{\partial h}{\partial
\gamma _{mn}},\quad \text{and}\quad \delta N=\dot{\varepsilon}(\eta )
\end{equation}
provided $\varepsilon (\eta )$ vanishes at the end points, $\varepsilon
(\eta _i)=\varepsilon (\eta _f)=0$. The invariance $\delta I=0$ holds for
any path $\gamma (\eta )$, $\pi (\eta )$ and not just for those paths at
which the action is stationary. In addition, as is evident in the action $%
J[\gamma ]$, there is an additional invariance under global ($\eta $%
-independent) ``conformal'' transformations, $\gamma _{ij}\rightarrow \psi
^4\gamma _{ij}$. The corresponding conserved quantity is $\func{tr}\pi .$ To
appreciate the significance of this conserved quantity note that 
\begin{equation}
\func{tr}\pi =\gamma ^{mn}\pi _{mn}=\frac{\gamma ^{mn}\dot{\gamma}_{mn}}{2N}%
=\frac 1{2N\gamma }\frac{d\gamma }{d\eta }=\frac 1{2\gamma }\frac{d\gamma }{%
d\tau }\,,
\end{equation}
so that the determinant $\gamma $ expands or contracts at a constant
relative rate. In particular, if the initial velocity happens to be such
that $\func{tr}\pi =0$, then $\gamma $ remains fixed at its constant initial
value.

\section{Geometrodynamics: the Ultralocal Case}

The system we study is a \emph{single} dust cloud. To the dust cloud we
associate a probability distribution $P$ given by a product of the
distributions, eq.(\ref{p(y|x)}) of the individual particles, 
\begin{eqnarray}
P[y|\gamma ] &=&\prod_xp\left( y(x)|x,\gamma _{ij}(x)\right)  \nonumber \\
&=&\left[ \prod_x\frac{\gamma ^{1/2}(x)}{(2\pi )^{3/2}}\right] \exp \left[
-\frac 12\sum_x\gamma _{ij}(x)(y^i-x^i)(y^j-x^j)\right] .
\label{cloud state}
\end{eqnarray}

It was the necessity to quantify whether we can distinguish a test particle
at $x$ from its neighbor at $x+dx$ that led us to introduce the metric $%
\gamma _{ij}$ in the first place. When we consider the change from an
earlier state $\gamma $ to a later state $\gamma +\Delta \gamma $ the
distinguishability problem manifests itself yet again. Even if we had
managed to distinguish a test particle at $x$ from a neighboring test
particle at $x+dx$, there is no guarantee that the particle that earlier had
coordinates $x$ will be the \emph{same} particle that will later be found at 
$x$. Particles do not just need to be identified, they need to be
re-identified. For the invisible points of space this difficulty is only
exacerbated because the re-identification of points depends on the state of
motion of the test particles. If we allow for the possibility of particles
moving past each other we conclude that the points of space cannot be
treated as enduring things. And this is precisely where the model discussed
in this section becomes unrealistic: we maintain such a strict
correspondence between a test particle and the point it occupies that we end
up treating the individual points of space as if they were real enduring
objects. A more realistic model of space should deal with \emph{several}
potentially coexisting dust clouds in relative motion.

Once a dust particle in the earlier state $\gamma $ is identified with the
label $x$, we will assume that this particle can be assigned the same label $%
x$ as it evolves into the later state $\gamma +\Delta \gamma $. These are
commoving coordinates. Then we can write the change $\Delta \ell $ between $%
P[y|\gamma +\Delta \gamma ]$ and $P[y|\gamma ]$, eq.(\ref{cloud state}),
from their relative entropy, 
\begin{equation}
S[\gamma +\Delta \gamma ,\gamma ]=-\int \left( \stackunder{x}{\Pi }%
dy(x)\right) P[y|\gamma +\Delta \gamma ]\log \frac{P[y|\gamma +\Delta \gamma
]}{P[y|\gamma ]}=-\frac 12\Delta \ell ^2
\end{equation}
Since $P[y|\gamma ]$ and $P[y|\gamma +\Delta \gamma ]$ are products $%
S[\gamma +\Delta \gamma ,\gamma ]$ can be written as a sum over the
individual particles, 
\begin{equation}
S[\gamma +\Delta \gamma ,\gamma ]=\sum_x\,S[\gamma (x)+\Delta \gamma
(x),\gamma (x)]=-\frac 12\sum_x\Delta \ell ^2(x)\,,
\end{equation}
where   
\begin{equation}
\Delta \ell ^2(x)=g^{ij\,kl}(x)\Delta \gamma _{ij}(x)\Delta \gamma
_{kl}(x)\,,
\end{equation}
with $g^{ij\,kl}$ given by eq.(\ref{metric g}). Therefore, the overall
change in going from $\gamma $ to $\gamma +\Delta \gamma $ is 
\begin{equation}
\Delta \ell ^2=\sum_x\Delta \ell ^2(x)=\int dx\,\rho (x)\Delta \ell ^2(x)\,,
\label{Delta lscr}
\end{equation}
where we have written the discrete sum as an integral -- the number of dust
particles within $dx$ is $dx\rho (x)$.

Having given a sufficient specification of what we mean by a state of the
system we can now proceed to formulate its ED. Once again we ask, `Given
initial and final states, what trajectory is the system expected to follow?'
and the answer follows from the implicit assumption that there exists a
continuous trajectory, but here we must pay closer attention to what
precisely we mean by `trajectory'. Indeed, if predicting changes is just a
matter of careful consistent manipulation of the available information, then
we must recognize that we know more than just that the product state eq.(\ref
{cloud state}) must evolve through a continuous sequence of intermediate
states. We also know that each and every one of the individual factors $%
p(y|x,\gamma )$ must also evolve continuously through a sequence of
intermediate states to reach the corresponding final state. This means that
instead of one parameter $\omega $ there are many such parameters, one for
each position $x$, and there is no reason why they should all take the same
value. In other words, the intermediate states $\gamma _{{}\omega }$ should
be labeled by a local function $\omega (x)$ rather than a single global
parameter $\omega $. A continuous sequence of states $\gamma _{{}\omega }$
interpolating between the initial $\gamma $ and the final $\gamma +\Delta
\gamma $ can be defined by imposing $\omega (x)=\zeta f(x)$ where $f(x)$ is
a fixed positive function and the parameter $\zeta $ varies from $0$ to $%
\infty $. There is no single trajectory; each choice of the function $f(x)$
defines one possible trajectory. In a sense, the cloud follows many
alternative paths ``simultaneously''. To guarantee consistency we should
check that physical predictions are independent of the choice of the
arbitrary function $f(x)$.

Before we formulate the ED we should remark on the significance of
invariance under choices of $f(x)$. The product state $P[y|\gamma ]$
provides the only definition of what an instant is, of which states $%
p(y|x^{\prime },\gamma ^{\prime })$ at distant points $x^{\prime }$ we can
agree to call simultaneous with a certain state $p(y|x,\gamma )$ at the
point $x$. Therefore, if there is no unique sequence of intermediate states,
then there is no unique, absolute definition of simultaneity. We see here a
kind of foliation invariance, a rudimentary, and yet extreme form of local
Lorentz invariance. Since the metric $\gamma _{{}\omega }$ of the
intermediate states $P[y|\gamma _{{}\omega }]$ remains positive for
arbitrary choices of the function $\omega (x)$ the analogues of the light
cones are collapsed into light lines. The invariant speed -- the speed of
light -- is zero. The GD model described here resembles the so-called
ultralocal or strong gravity theories \cite{Isham76}--\cite{Pilati82} more
closely than it resembles general relativity.

Now we address the question: Given initial and final states, $\gamma $ and $%
\gamma +\Delta \gamma $, what are the possible trajectories? Let $\eta $ be
an arbitrary time parameter labeling successive intermediate states. The
initial state is $\gamma _{ij}(\eta ,x)=\gamma _{ij}(x)$, the final state is 
$\gamma _{ij}(\eta +\Delta \eta ,x)=\gamma _{ij}(x)+\Delta \gamma _{ij}(x)$,
and the intermediate states are $\gamma _{ij}(\eta +d\eta ,x)=\gamma
_{ij}(x)+d\gamma _{ij}(x)$. To determine the intermediate state $\gamma
+d\gamma $ one varies over $d\gamma _{ij}$ to maximize the entropy 
\begin{equation}
S[\gamma +d\gamma ,\gamma ]=-\int \left( \stackunder{x}{\Pi }dy(x)\right)
P[y|\gamma +d\gamma ]\log \frac{P[y|\gamma +d\gamma ]}{P[y|\gamma ]}=-\frac
12d\ell ^2\,,
\end{equation}
where 
\begin{equation}
d\ell ^2=\int dx\,\rho (x)d\ell ^2(x)\quad \text{with}\quad d\ell
^2(x)=g^{ij\,kl}(x)d\gamma _{ij}(x)d\gamma _{kl}(x)\,,
\end{equation}
subject to independent constraints at each point $x$, 
\begin{equation}
d\ell _f(x)=\omega (x)d\ell (x)
\end{equation}
where 
\begin{equation}
d\ell _f^2(x)=g^{ij\,kl}(x)\left( \Delta \gamma _{ij}(x)-d\gamma
_{ij}(x)\right) \left( \Delta \gamma _{kl}(x)-d\gamma _{kl}(x)\right) 
\end{equation}
Introducing Lagrange multipliers $\lambda (x)/2$, 
\begin{equation}
0=\delta \left[ \int dx\,\rho (x)\left\{ -\frac 12d\ell ^2(x)+\frac{\lambda
(x)}2\left( \omega ^2(x)d\ell ^2(x)-d\ell _f^2(x)\right) \right\} \right] 
\end{equation}
the result, $d\gamma _{ij}(x)=\chi (x)\Delta \gamma _{ij}(x)$, coincides
with the single point result, eq.(\ref{single point dgamma}) for each value
of $x$. Substituting $d\gamma _{ij}$ into $d\ell (x)$ and $d\ell _f(x)$ we
get $d\ell (x)=\chi \Delta \ell (x)$ and $d\ell _f(x)=[1-\chi (x)]\Delta
\ell (x)$, so that 
\begin{equation}
d\ell (x)+d\ell _f(x)=\Delta \ell (x)\,.
\end{equation}
The conclusion is that the states of the individual particles evolve
independently of each other along geodesics in the single point
configuration space given by eqs.(\ref{single point motion c}-\ref{single
point motion d}). The dynamics of the cloud is independent of the choice of $%
\omega (x)$ as desired -- this is foliation invariance.

The ultralocal statistical GD deduced in the previous paragraphs is the
dynamics of a large or perhaps infinite number of independent subsystems.
The action for the whole cloud can be written as the sum of the individual
particle actions given in eq.(\ref{single point J}). Thus, the proposed
action is 
\begin{equation}
J[\gamma ,\dot{\gamma}]=\int_{\eta _i}^{\eta _f}d\eta \,\int dx\,\rho \left(
g^{ij\,kl}\dot{\gamma}_{ij}\dot{\gamma}_{kl}\right) ^{1/2},
\label{ultralocal J}
\end{equation}
where $\dot{\gamma}_{ij}=\partial \gamma _{ij}/\partial \eta $. In commoving
coordinates $\dot{\rho}=\partial \rho /\partial \eta =0$. It is
straightforward to develop the constrained Hamiltonian formalism and recover
the single particle equations of motion.

Notice that the actual distance from the initial state to the final state
along a certain path is given by eq.(\ref{Delta lscr}), 
\begin{equation}
\ell =\int_i^fd\eta \,\left( \dot{\ell}^2\right) ^{1/2}=\int_{\eta _i}^{\eta
_f}d\eta \left[ \int dx\,\rho \,g^{ij\,kl}\dot{\gamma}_{ij}\dot{\gamma}%
_{kl}\right] ^{1/2}.
\end{equation}
Therefore, unlike the action for a single point, eq.(\ref{single point J}),
the action (\ref{ultralocal J}) is not the natural arc length. The dust
cloud does not evolve along a geodesic. The reason for this can be traced to
the additional constraint that individual particles evolve continuously,
which allows a multitude of different trajectories and leads to foliation
invariance.

\section{Conclusions}

One idea explored in this work is whether it is possible to establish a
connection between ordinary spatial distances and the information metric of
Fisher and Rao -- whether one can explain the notion of spatial distance. We
succeeded in describing the information geometry that derives from
considerations of distinguishability among particles; particles that are
easily confused are said to be near, those that are easily distinguished are
farther apart. The idea is that distances between particles are not
distances between structureless points but distances between probability
distributions.

According to Euclid, a point is that which has no size. General relativity
was founded upon a revision of Euclid's fifth postulate. Statistical
geometrodynamics is founded upon the further revision of Euclid's first
definition, the notion of structureless points.

The second idea we explored is whether Einsteinian macroscopic
geometrodynamics is derivable from an underlying microscopic statistical
theory purely on the basis of principles of inference, without additional
postulates of a more ``physical'' nature. We can only claim a partial
success; the result is close enough to be promising. The model GD we
obtained satisfies the main requirement, it describes the dynamics of a
geometry; it is related to gravity because it describes an ultralocal
gravity theory; and it exhibits foliation invariance. Moreover, the somewhat
puzzling fact that space and time are so different and yet enter the
formalism in such a symmetric way receives a natural explanation: a time
interval refers to the extent we can distinguish an earlier state from a
later state of the same system, while a spatial distance refers to the
extent we can distinguish two different systems.

Einstein's GD might be recovered by making a different choice of the states
and variables that describe the gravitational degrees of freedom. Two
possible alternative choices were suggested. First, one should avoid a too
strict correspondence between a test particle and the point it occupies
because this treats the individual points of space as if they were real
objects. Second, it may be that the Fisher-Rao metric does not describe the
full geometry of space, as we assumed in this work, but only describes its
conformal geometry.

Should the ideas proposed here prove successful one can further expect that
the currently popular approaches to a quantum theory of gravity will require
revision.

\end{document}